\documentclass[twocolumn,showpacs,preprintnumbers,amsmath,amssymb]{revtex4}
\pdfoutput=1

\usepackage{xspace}
\usepackage{graphicx}

\def\beq{\begin{equation}}
\def\eeq{\end{equation}}

\def\bea{\begin{eqnarray}}
\def\eea{\end{eqnarray}}

\newcommand{\roughly}[1]%
    {{\mathrel{\raise.3ex\hbox{$#1$\kern-.75em\lower1ex\hbox{$\sim$}}}}}

\newcommand{\gsim}{\mathrel{\roughly>}}

\newcommand{\scr}[1]{\ensuremath{\mathcal{#1}}}

\newcommand{\be}{\ensuremath{\beta}}

\newcommand{\De}{\ensuremath{\Delta}}

\newcommand{\la}{\ensuremath{\lambda}}
\newcommand{\La}{\ensuremath{\Lambda}}

\newcommand{\Si}{\ensuremath{\Sigma}}

\newcommand{\hc}{\ensuremath{\mbox{h.c.}}}

\newcommand{\tr}{\mathop{\mbox{tr}}}

\newcommand{\avg}[1]{\ensuremath{\langle{#1}\rangle}}

\newcommand{\GeV}{\ensuremath{\mathrm{~GeV}}}
\newcommand{\TeV}{\ensuremath{\mathrm{~TeV}}}

\newcommand{\Eq}[1]{Eq.~(\ref{#1})}

\newcommand{\Ref}[1]{Ref.~\cite{#1}}
\newcommand{\Refs}[1]{Refs.~\cite{#1}}

\begin{document}


\title{Superconformal Technicolor}

\author{Aleksandr Azatov}
\email{aleksandr.azatov@roma1.infn.it}
\author{Jamison Galloway}
\email{jamison.galloway@roma1.infn.it}
\affiliation{Dipartimento di Fisica, Universit\`a di Roma ``La Sapienza"\\
and INFN Sezione di Roma, I-00185 Roma}
\author{Markus A. Luty}%
\email{luty@physics.ucdavis.edu}
\affiliation{%
Physics Department,~University of California Davis\\
Davis,~California 95616}%


\begin{abstract}
In supersymmetric theories with a strong conformal sector,
soft supersymmetry breaking at the TeV scale naturally gives
rise to confinement and chiral symmetry breaking at the same scale.
We investigate models where such a sector dynamically
breaks electroweak symmetry.
We consider two scenarios, one where the strong dynamics induces
vacuum expectation values for elementary Higgs fields,
and another where the strong dynamics is solely responsible
for electroweak symmetry breaking. 
In both cases there is no fine tuning required to explain
the absence of a Higgs boson below the LEP bound,
solving the supersymmetry naturalness problem.
A  good precision electroweak fit can be obtained,
and quark and lepton masses are generated without flavor-changing
neutral currents.
Electroweak symmetry breaking may be dominated either by the
elementary Higgs bosons or by the strong dynamics.
In addition to standard superymmetry collider signals, 
these models predict production of multiple heavy standard
model particles ($t$, $W$, $Z$, and $b$) from decays of
resonances in the strong sector.
\end{abstract}


\maketitle

{\it Introduction---}Supersymmetry (SUSY) is widely considered to be the
most plausible framework for physics beyond the standard model
of particle physics. 
It offers an elegant explanation of the fact that electroweak breaking scale
$\sim 100\GeV$ is much smaller than the Planck scale $\sim 10^{19}\GeV$,
without fine tuning fundamental parameters.
The minimal supersymmetric standard model (MSSM)
also contains a viable dark matter candidate
and gives a calculable framework for addressing other
fundamental issues in particle physics and cosmology.
However, there is a serious problem with electroweak symmetry breaking in
the MSSM: the lightest Higgs has a mass that is generically
$m_h < m_Z \simeq 90\GeV$,
while the experimental bound from LEP is $m_h > 115$~GeV \cite{LEPHiggs}.
The Higgs mass can be raised at the cost of re-introducing tuning
at the $1\%$ level, or by extending the model in various ways \cite{SUSYHiggs}.
In this Letter, we propose to solve this problem by combining
supersymmetry with strong dynamics at the TeV scale.
A companion paper \cite{SCTC2} gives many additional details.

The electroweak scale in the MSSM is determined by the scale of
soft SUSY breaking.
We assume that in addition there is a strongly-coupled sector of the
theory with 
conformal (scale) invariance.
An example of such a sector is SUSY QCD with
$N_f \simeq 2 N_c$ \cite{Seiberg}.
Soft SUSY breaking in the strong sector also softly breaks the conformal
invariance.
SUSY breaking in the strong sector gives mass to all scalars
(since only unbroken SUSY can forbid these masses),
while fermions generally remain massless
due to unbroken chiral symmetries.
It is therefore very plausible that the dynamics of 
SUSY QCD at the SUSY breaking scale is qualitatively similar
to non-SUSY QCD, with confinement and chiral symmetry breaking in particular.
Since the coupling is already strong at the SUSY breaking scale,
these effects occur at this scale.
In such models the strong sector can dynamically break electroweak
symmetry, as in technicolor models \cite{TC}.
Since the scale of dynamical electroweak symmetry breaking is determined
by the soft breaking of conformal symmetry, this is a SUSY version of
conformal technicolor \cite{CTC}, so we refer to it as superconformal
technicolor \cite{sannino}.
We assume that the SUSY breaking scale is the same order of
magnitude in the MSSM and the strong sector, which is natural in 
many models of SUSY breaking.
This class of models therefore gives a plausible framework
for supersymmetry \emph{and} strong dynamics at the same scale.
In \Ref{CTCflavor} this mechanism was employed with a SUSY breaking
scale above the electroweak scale to give a realistic model for
flavor in conformal technicolor (the pioneering work in this direction
is \Ref{bosonicTC}).
In the present work, we investigate SUSY breaking 
and strong dynamics at the TeV scale.
Early attempts in this direction posited dynamical SUSY breaking at the
TeV scale \cite{supercolor}, but this is problematic for both theoretical
and phenomenological reasons.
A realistic model was constructed in \Ref{LTG}.
The present work improves on that work in giving a general and robust
mechanism for the coincidence of the scales of SUSY breaking
and strong dynamics.

{\it Induced electroweak symmetry breaking---}In these models
there are two potential sources of electroweak
symmetry breaking, the strong sector and the elementary Higgs fields
of the MSSM.
We first consider a scenario where electroweak symmetry breaking is 
induced by the strong sector, but the $W$ and $Z$ masses are dominated
by the contribution from the elementary Higgs fields.
A minimal strong sector has fields transforming under
$SU(2)_{\rm SC} \times SU(2)_W \times U(1)_Y$
as
\beq
\Psi \sim (2, 2)_0,
\quad
\tilde{\Psi}_1 \sim (2, 1)_{\frac 12},
\quad
\tilde{\Psi}_2 \sim (2, 1)_{-\frac 12},
\eeq
plus 2 copies of $(2, 1)_{\frac 12}
\oplus (2, 1)_{-\frac 12}$ fields that play no role
in breaking electroweak symmetry.
(The hypercharge assignments ensure that there are
no fractionally charged states in the strong sector.)
The fields $\Psi$ and $\tilde\Psi$ have the quantum numbers
of the technifermions of minimal technicolor \cite{TC}.
The soft SUSY breaking terms explicitly break
the global symmetry of the strong sector to
$SU(2)_L \times SU(2)_R$.
SUSY breaking in the strong sector is assumed to trigger
confinement 
and chiral symmetry breaking by a fermion condensate
$\avg{\Psi \tilde{\Psi}} \ne 0$, as in technicolor.
(It is also natural to have a larger group of approximate
symmetries due to special structure of the soft SUSY breaking terms,
in which case there will be additional light pseudo-Nambu Goldstone
bosons.)
Stabilizing runaway directions in the strong sector
requires additional interactions, which are discussed in
\Ref{SCTC2}.
The strong sector is
coupled to the MSSM Higgs fields via the superpotential couplings
\beq
\label{WHstrong}
W = \la_u H_u (\Psi \tilde{\Psi}_2) + \la_d H_d (\Psi \tilde{\Psi}_1).
\eeq
The operators $\Psi\tilde\Psi$ have dimension $\simeq \frac 32$
above the SUSY breaking scale, and so the couplings $\la_{u,d}$ have 
mass dimension $\simeq +\frac 12$.
We require that the couplings $\la_{u,d}$ be large enough to
be important at the SUSY breaking scale, but not non-perturbatively
large.
This is a coincidence of scales between a relevant SUSY preserving
coupling and the SUSY breaking scale, similar to the problem of
why the superpotential term $\mu H_u H_d$ has a coupling
$\mu \sim 100\GeV$.
The simplest solution to the ``$\mu$ problem'' is the 
Giudice-Masiero mechanism \cite{GM}, and 
\Ref{SCTC2} gives a generalization of this mechanism that
can explain the required values of $\la_{u,d}$.


We assume that the strong sector dynamically breaks electroweak symmetry
with order parameter $f$ somewhat below what is required to explain
the $W$ and $Z$ masses, {\it e.g.\/}~$f \simeq 100\GeV$.
We expect that the strong sector contains massive ``hadron'' states
at a scale $\La \sim 4\pi f \sim \mbox{TeV}$.
We assume that the elementary Higgs fields $H_{u,d}$ have masses below
$\La$, so the effective theory below the scale $\La$
contains these fields.
%
The $SU(2)_L \times SU(2)_R$ symmetry is nonlinearly
realized in the low-energy effective theory
by $\Si(x) \in SU(2)$ transforming 
as $\Si \mapsto L \Si R^\dagger$.
The elementary Higgs fields and couplings in 
\Eq{WHstrong} can be combined into a
spurion 
\beq
\scr{H} \la = 
\begin{pmatrix}
\la_d H_d^0 & \la_u H_u^- \cr
\la_d H_d^+ & \la_u H_u^0 \cr
\end{pmatrix}
\mapsto L \scr{H} \la R^\dagger.
\eeq
The strong dynamics generates
new contributions to the Higgs potential
\beq
\label{Vtad}
\De V_{\rm eff} = \frac{\La^4}{16\pi^2}
\left[ \frac{c_1}{\La} \tr(\Si^\dagger  \scr{H} \la) + \hc
+ \scr{O}\left( \frac{ \scr{H} \la}{\La} \right)^2 \right].
\eeq
with $c_1 \sim 1$ \cite{NDA}.
This contains a linear term 
for the Higgs fields, so the Higgs fields get VEVs 
even for $m_{H_{u,d}}^2 > 0$, which we assume to be the case.
(In standard SUSY scenarios $m_{H_{u,d}}^2 > 0$ at high scales
and renormalization group running results in
$m_{H_u}^2 < 0$ at the TeV,
but more general boundary conditions at high scales can lead
to $m_{H_{u,d}}^2 > 0$ at the TeV scale.)
We assume that this generates VEVs for the elementary Higgs
fields with $v_u^2 + v_d^2 \gg f^2$
({\it e.g.\/}\ for $f \simeq 100\GeV$,
$\sqrt{v_u^2 + v_d^2} \simeq 225\GeV$).
The higher order terms in \Eq{Vtad} are negligible if
$m_{H_{u,d}} \ll 4\pi f^2 / v$.
Note that we get a stable minimum even if we neglect the
quartic terms in the potential and $B\mu$ term, and the physical Higgs
masses are given by the quadratic terms in the potential
in this limit.
Including the full potential, the predictions are more complicated,
but the Higgs masses are still arbitrary parameters depending
on the SUSY breaking mass terms.
The physical Higgs masses can have any value
$\sim 100\GeV$ without fine tuning,
so this completely solves the SUSY Higgs mass problem.

The quark and lepton masses arise from conventional Yukawa 
couplings to $H_{u,d}$,
which have a minimal flavor-violating structure.
Since $\avg{H_{u,d}}$ is the dominant
source of electroweak symmetry breaking, the Yukawa
couplings are perturbative, even for the top quark.
Therefore, there is no flavor problem associated with the
strong dynamics.

\begin{figure}[hbt]
\begin{center}
\includegraphics[width=8.0cm]{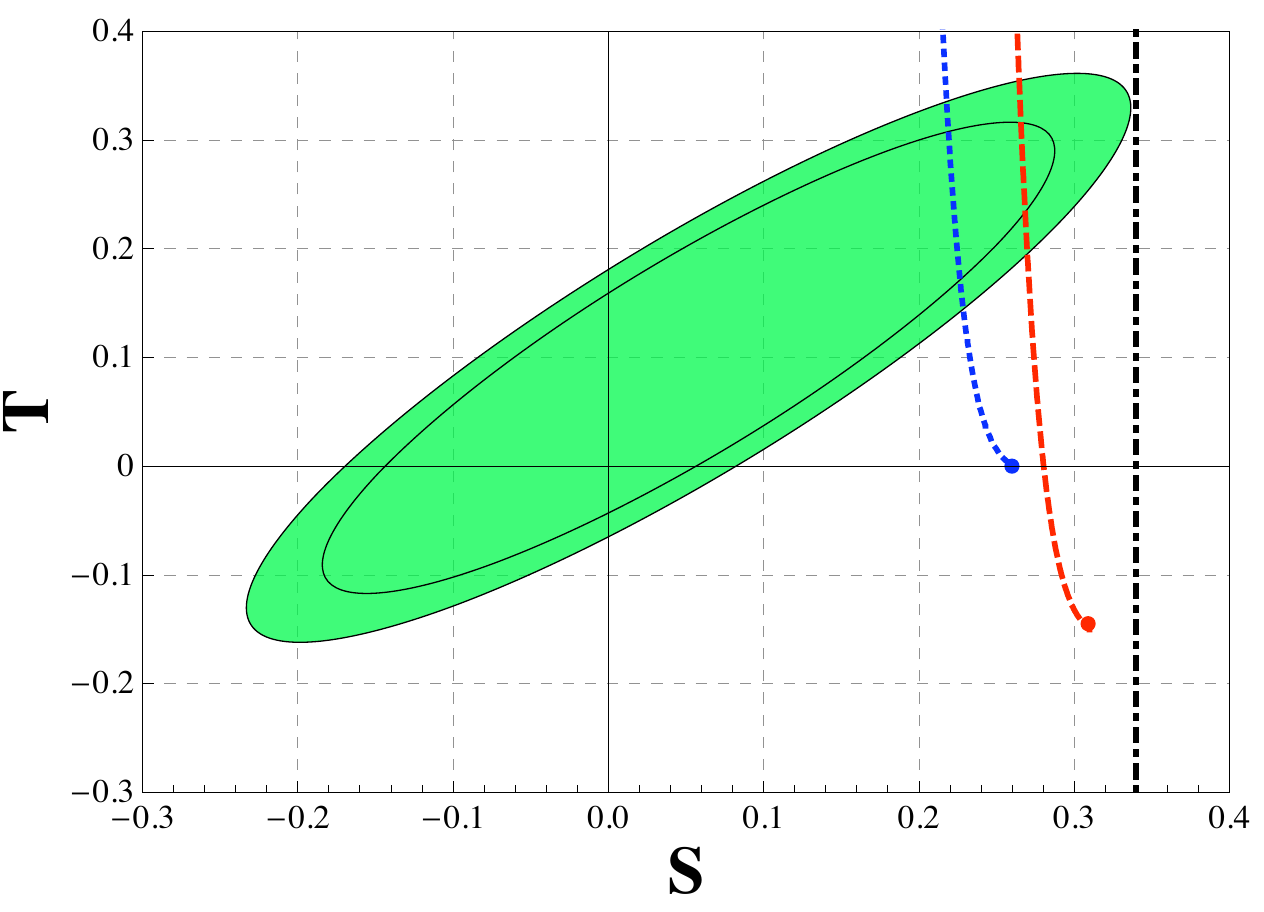}
\caption{Precision electroweak fit.
The inner (outer) ellipse is the $95\%$ ($99\%$) confidence level
allowed region in the $S$, $T$ plane
with reference Higgs mass $120\GeV$ \cite{Gfitter}.
The dotted blue (dashed red) line corresponds to a model with
a light elementary Higgs at 130 (350) GeV,
with $f=100\, {\rm GeV}$, $\tan \beta = 2$, and $B\mu = 0$.
The lines end when $\la_u v_u \simeq \lambda_d v_d$,
where $T$ is dominated by the light Higgs contribution.
The dot-dashed black line is for the model with no light Higgs.
The plot assumes that the UV contribution to the $S$ parameter
is given by the QCD value, while the UV contribution to the
$T$ parameter is estimated using NDA.
Both of these have large theoretical uncertainties, so this plot is
meant only to be suggestive.}
\label{fig:ST}
\end{center}
\end{figure}

We now turn to the phenomenology of this model.
Early work on technicolor theories with Higgs
scalars can be found in \Refs{TCscalar}.
We first discuss the precision electroweak fit.
The strong sector has $N_c = 2$ and only one weak doublet,
so the contributions to the $S$ and $T$ parameters
from the strong sector are not dangerously large to begin
with, and there are large theoretical uncertainties in
their values.
In fact, general theoretical arguments suggest
that the $S$ parameter is suppressed in theories that are
conformal above the chiral symmetry breaking scale \cite{SundrumHsu}.
Recent lattice simulations give some support for this
behavior \cite{Slattice}.
In the present model the IR contribution to $S$ from the strong
sector is reduced compared
to a conventional technicolor theory
because the PNGBs are heavy, and because there is a light Higgs in the
spectrum. 
%
Custodial symmetry can be broken in the strong sector by
$\la_u v_u \ne \la_d v_d$.
We assume that this contribution to the $T$ parameter
is positive, as suggested by perturbation theory.
This means that the theory has an adjustable parameter that allows
a good precision electroweak fit (similar to the Higgs mass in the
standard model).
We can easily obtain a good precision electroweak
fit, even if we assume (pessimistically)
that the UV contribution to the $S$ parameter
has the value obtained by extrapolation from QCD \cite{SQCD}. 
This is illustrated in Fig.~\ref{fig:ST}.

Another important precision electroweak constraint is the coupling of the
$Z$  to left-handed $b$ quarks.
In this model the leading correction enters at
$\mathcal O(y^2 \lambda^2)$, with $y$ the Yukawa coupling to SM quark fields.
The coupling $g_{Z\bar b  b}$ agrees with the standard model at the $0.25\%$
level, which gives a constraint $v < 5.6 f$.
This is easily satisfied given the other constraints we have considered, and 
we conclude that this coupling does not significantly restrict the
viable parameter space.

We now discuss the signals at the LHC.
In addition to the standard MSSM signatures, the theory has
new signatures from the extended Higgs sector.
In the simplified limit discussed above,
the $CP$ even scalars have masses $m_{H_{u,d}}$ 
while for the $CP$ odd scalars we have (for $f \ll v$)
\begin{align}
m_{A^0_1}^2 &= m_{H^\pm_1} = \frac{m_{H_u}^2 m_{H_d}^2}
{m_{H_u}^2 \sin^2\be + m_{H_d}^2 \cos^2\be},
\\
m_{A^0_2}^2 &= m_{H^\pm_2} = \frac{v^2}{f^2}
(m_{H_u}^2 \sin^2\be + m_{H_d}^2 \cos^2\be),
\end{align}
where $\tan\be = v_u / v_d$.
The heavier 
mass eigenstates $A_2^0$ and $H^\pm_2$ are 
dominantly PNGBs from the strong sector,
with mixing of order $f/v$ with the elementary Higgs fields.
%
%
%
The $A_2^0$ can be singly produced by gluon fusion via
a top loop, with a rate suppressed by $f^2/v^2$.
$A_2^0$ and $H_2^\pm$ can also be pair produced via
heavy resonances in the strong sector.
Dominant decay modes are
$A_2^0 \to \bar{t}t, W^\pm H^\mp , Z h^0, A_1^0h^0$
and $H_2^+ \to \bar{b}t , W^+ h^0, H_1^+ h^0$.
The $h^0$ decays dominantly to $\bar{b}b$ or $WW$/$ZZ$
depending on its mass, so this leads to
events with multiple heavy standard model particles
($W$, $Z$, $t$ and/or $b$).
Another signal is resonances in the strong sector
with masses of order $4\pi f \sim \mbox{TeV}$.
Analogy with QCD suggests that the theory may have a prominent
isotriplet vector resonance, the $\rho_T$.
This can be singly produced via mixing with the $W$ and $Z$ of
order $g/4\pi$, or via weak boson fusion.
The $\rho_T$ will generally have strong decays to pairs of PNGBs, but
because of the large elementary Higgs VEVs, the $A_2^0$ and $H_2^\pm$
masses can be sufficiently large that decays to these states are
kinematically forbidden.
The effective field theory expansion breaks down in this
regime, but we still expect it to be qualitatively reliable.
In this case the $\rho_T$ will be a narrow resonance, similar
to a $W'$ and $Z'$.
Techniscalars charged under $SU(2)_L$ and $SU(2)_R$
generally have different SUSY breaking masses,
so there need not be any approximate 
symmetry that interchanges $SU(2)_L$ and $SU(2)_R$,
analogous to parity in QCD.
This means that $\rho_T$ can decay to either $WW$ and $WWW$.
The $\rho_T$ can also decay via mixing with the $W$ and $Z$.

{\it Strong electroweak symmetry breaking---}We now consider
another scenario
where there are no elementary Higgs fields below the TeV scale,
and electroweak symmetry is broken entirely by the strong sector.
This arises in a different parameter regime of the model
described above, as follows.
We assume that the couplings $\la_{u,d}$ in \Eq{WHstrong} get strong
at a scale $\La_* > \mbox{TeV}$.
Results on non-perturbative dynamics of SUSY gauge theories 
\cite{Seiberg} indicate
that below the scale $\La_*$ the
theory flows to a new fixed point where these couplings are
strong.
In this new fixed point, $H_{u,d}$ become operators of the
strong sector with dimension $\simeq \frac 32$.
This means that the Yukawa couplings of $H_{u,d}$ to 
quarks and leptons become irrelevant interactions below the scale $\La_*$,
scaling as $(E/\La_*)^{1/2}$.
In order to avoid too much suppression for the top quark mass,
we cannot have $\La_*$ arbitrarily far above the TeV scale.
If $\La_* \gg \mbox{TeV}$ the top quark Yukawa coupling
gets strong at some scale above $\La_*$, indicating top quark
compositeness at high scales.
Alternatively, models with $\La_* \sim \mbox{TeV}$ are natural
with a mechanism to explain the coincidence of scales,
as described above.
For $\La_* \gsim \mbox{TeV}$,
quark and lepton masses arise from irrelevant
interactions at the TeV scale, as in technicolor.
However, these interactions originate from
Yukawa couplings with minimal flavor violation, and there is no flavor
problem associated with the strong breaking of electroweak symmetry.

At the TeV scale, soft SUSY breaking in the strong sector is assumed
to trigger confinement and electroweak symmetry breaking, as discussed
above.
The soft SUSY breaking terms can be chosen so that the strong sector
has a minimal symmetry breaking structure
$SU(2)_L \times SU(2)_R \to SU(2)$, so the only strong degrees of
freedom below the TeV scale are the longitudinal components of the
$W$ and $Z$.
The spectrum at the TeV scale therefore includes all of the MSSM
fields minus the Higgs sector, with strong resonances at the
scale $4\pi v \sim 3\TeV$.

A good precision electroweak fit can be obtained with the help
of a $T$ parameter induced by $\la_u \ne \la_d$.
Assuming that the $S$ parameter is given by the QCD value,
the precision electroweak fit is shown in Fig.~\ref{fig:ST}.
A good fit can be obtained if the UV contribution
to the $S$ parameter is reduced compared to this estimate
(as we expect, as discussed above), and
the contribution to the $T$ parameter from $\la_u \ne \la_d$
is positive (as expected from perturbation theory).  
The correction to $g_{Z\bar{b}b}$ is of order $0.8\%$,
with large theoretical uncertainties.
This is roughly 3 times the experimental precision
so there is some tension, but given the large uncertainties 
this does not rule out the model.
The collider phenomenology consists of SUSY signals, plus technicolor
resonances at the $3\TeV$ scale.
The $\rho_T$ can decay to both $WW$ and $WWW$ as
described above, which distinguishes it from the conventional
technirho.

{\it Conclusions---}We have described models that solve the
SUSY Higgs mass problem via strong dynamics at the TeV scale.
The models consist of the MSSM plus a sector with a strong conformal
fixed point.
In such models, it is natural for the strong sector
to dynamically break electroweak
symmetry at the soft SUSY breaking scale.
We considered two scenarios, one in which the strong breaking of
electroweak symmetry induces the elementary Higgs VEVs, and one
in which strong electroweak symmmetry breaking dominates.
In both scenarios the experimental bounds on light Higgs bosons
are easily satisfied without tuning, and no additional flavor
problem is introduced.
Both scenarios have a dark matter candidate.
However, gauge coupling unification is no longer a prediction of
the minimal model described here,
since the strong sector affects the evolution
of the $SU(2)_W \times U(1)_Y$ gauge couplings but not
$SU(3)_C$.
Unification can be accommodated with additional matter fields,
which however have no other apparent motivation in this framework.
In conclusion,
we believe that this is a plausible framework for electroweak
symmetry breaking, and that the new signals suggested by these
models deserve additional investigation.

{\it Acknowledgements---}We thank R. Contino, R. Kitano, T. Okui, 
and J. Terning for discussions.
We also thank J. Serra for collaboration at early stages of this work.
M.A.L. was supported by DOE grant
DE-FG02-91-ER40674.


\begin{thebibliography}{99}

\bibitem{LEPHiggs}
R.~Barate {\it et al.} [LEP Higgs Working Group],
Phys.\ Lett.\  {\bf B565}, 61-75 (2003) [hep-ex/0306033].

\bibitem{SUSYHiggs}
For an entry into the literature, see
  A.~Birkedal, Z.~Chacko, Y.~Nomura,
  Phys.\ Rev.\  {\bf D71}, 015006 (2005)
  [hep-ph/0408329];
  A.~Maloney, A.~Pierce, J.~G.~Wacker,
  JHEP {\bf 0606}, 034 (2006)
  [hep-ph/0409127];
  S.~Chang, C.~Kilic, R.~Mahbubani,
  Phys.\ Rev.\  {\bf D71}, 015003 (2005)
  [hep-ph/0405267];
S.~Chang, R.~Dermisek, J.~F.~Gunion, N.~Weiner,
  Ann.\ Rev.\ Nucl.\ Part.\ Sci.\  {\bf 58}, 75-98 (2008)
  [arXiv:0801.4554 [hep-ph]].

\bibitem{SCTC2}
  A.~Azatov, J.~Galloway, M.~A.~Luty,
  arXiv:1106.4815 [hep-ph].

\bibitem{Seiberg}
  N.~Seiberg,
  Nucl.\ Phys.\  B {\bf 435}, 129 (1995)
  [arXiv:hep-th/9411149].
  
\bibitem{TC}
For a review, see
  C.~T.~Hill, E.~H.~Simmons,
  Phys.\ Rept.\  {\bf 381}, 235-402 (2003)
  [hep-ph/0203079].

\bibitem{CTC}
  M.~A.~Luty, T.~Okui,
  JHEP {\bf 0609}, 070 (2006)
  [hep-ph/0409274];
  M.~A.~Luty,
  JHEP {\bf 0904}, 050 (2009)
  [arXiv:0806.1235 [hep-ph]].
  
  
\bibitem{sannino}
This name has also been used in for models that
do not use the conformal technicolor mechanism to break electroweak
symmetry in 
M.~Antola, S.~Di Chiara, F.~Sannino and K.~Tuominen,
  arXiv:1001.2040 [hep-ph] and
  arXiv:1009.1624 [hep-ph].




\bibitem{CTCflavor}
  J.~A.~Evans, J.~Galloway, M.~A.~Luty, R.~A.~Tacchi,
  JHEP {\bf 1104}, 003 (2011)
  [arXiv:1012.4808 [hep-ph]].
  
  
\bibitem{bosonicTC}
S.~Samuel,
  Nucl.\ Phys.\  B {\bf 347}, 625 (1990);
M.~Dine, A.~Kagan and S.~Samuel,
  Phys.\ Lett.\  B {\bf 243}, 250 (1990).
  
\bibitem{supercolor}
  S.~Dimopoulos and S.~Raby,
  Nucl.\ Phys.\  B {\bf 192}, 353 (1981);
  M.~Dine, W.~Fischler and M.~Srednicki,
  Nucl.\ Phys.\  B {\bf 189}, 575 (1981).

\bibitem{LTG}
  M.~A.~Luty, J.~Terning and A.~K.~Grant,
  Phys.\ Rev.\  D {\bf 63}, 075001 (2001)
  [arXiv:hep-ph/0006224].


\bibitem{GM}
  G.~F.~Giudice, A.~Masiero,
  Phys.\ Lett.\  {\bf B206}, 480-484 (1988).
  
\bibitem{NDA}
  H.~Georgi,
  Phys.\ Lett.\  {\bf B298}, 187-189 (1993)
  [hep-ph/9207278].

\bibitem{TCscalar}
  E.~H.~Simmons,
  Nucl.\ Phys.\  B {\bf 312}, 253 (1989);
  C.~D.~Carone and H.~Georgi,
  Phys.\ Rev.\  D {\bf 49}, 1427 (1994)
  [arXiv:hep-ph/9308205].
  
\bibitem{SundrumHsu}
  R.~Sundrum, S.~D.~H.~Hsu,
  Nucl.\ Phys.\  {\bf B391}, 127-146 (1993)
  [hep-ph/9206225].




\bibitem{Slattice}
  T.~Appelquist {\it et al.} [LSD Collaboration],
  [arXiv:1009.5967 [hep-ph]].


  



\bibitem{SQCD}
  M.~E.~Peskin, T.~Takeuchi,
  Phys.\ Rev.\ Lett.\  {\bf 65}, 964-967 (1990);
  B.~Holdom, J.~Terning,
  Phys.\ Lett.\  {\bf B247}, 88-92 (1990).

\bibitem{Gfitter}
  H.~Flacher, M.~Goebel, J.~Haller, A.~Hocker, K.~Monig, J.~Stelzer,
  Eur.\ Phys.\ J.\  {\bf C60}, 543-583 (2009).
  [arXiv:0811.0009 [hep-ph]].
  
  
\end{thebibliography}
\end{document}